\documentclass[final,11pt]{elsarticle}

\usepackage{verbatim,a4wide}

\usepackage{amsmath}
\usepackage{amssymb}
\usepackage{amsthm}
\usepackage[a4paper,left=2.5cm,right=2.5cm,top=2cm,bottom=3cm]{geometry}
\usepackage{graphicx}
\usepackage{caption}
\usepackage{subfig}
\usepackage{grffile}
\usepackage{float}
\usepackage{bbm}
\usepackage{indentfirst}
\usepackage{booktabs}
\usepackage{multirow}
\usepackage{rotating}
\usepackage{hyperref}
\usepackage{epstopdf}

\usepackage[cp1250]{inputenc}
\usepackage[OT4]{fontenc}
\usepackage[english]{babel}

\numberwithin{equation}{section}

\newtheorem{theorem}{Theorem}[section]

\newtheorem{prob}[theorem]{Problem}

\newdefinition{remark}[theorem]{Remark}
\newdefinition{example}[theorem]{Example}

\newcommand{\ra}[1]{\renewcommand{\arraystretch}{#1}}

\newcommand{\R}{\mathbb R}

\newcommand{\der}{\mbox{${\rm\,d}$}}
\newcommand{\dt}{\mbox{${\rm\,d}t$}}
\newcommand{\dv}{\mbox{${\rm\,d}v$}}

\newcommand{\rbinom}[2]{\mbox{$\displaystyle\binom{#1}{#2}^{\!\!-1}\!\!$}}

\journal{Journal of Computational and Applied Mathematics}

\begin{document}

\begin{frontmatter}

\title{Merging of B\'{e}zier curves with box constraints}

\author{Przemys{\l}aw Gospodarczyk\corref{cor}}
\ead{pgo@ii.uni.wroc.pl}
\author{Pawe{\l} Wo\'{z}ny}
\ead{Pawel.Wozny@ii.uni.wroc.pl}
\cortext[cor]{Corresponding author. Fax {+}48 71 3757801}
\address{Institute of Computer Science, University of Wroc{\l}aw,
         ul.~Joliot-Curie 15, 50-383 Wroc{\l}aw, Poland}

\begin{abstract}
In this paper, we present a novel approach to the problem of merging of B\'ezier curves with respect to the $L_2$-norm.
We give illustrative examples to show that the solution of the conventional merging problem may not be suitable
for further modification and applications. As in the case of the degree reduction problem,
we apply the so-called restricted area approach -- proposed recently in (P. Gospodarczyk, Computer-Aided Design 62 (2015), 143--151) -- to avoid certain defects and make the resulting curve more useful. A method of solving the new problem is based on box-constrained quadratic programming approach.
\end{abstract}


\end{frontmatter}

\section{Introduction}\label{Sec:intro}

Nowadays, people of various professions use different CAD systems. There are many ways to represent curves and surfaces,
therefore, the exchange of geometric data between those systems often requires approximate conversion.
As it was stated in~\cite{Hos87}, there are two main operations that should be considered: degree reduction and merging.
In the past $30$ years, both problems have been extensively investigated. In this paper, we focus on the constrained merging of segments of a composite B\'ezier curve, i.e., we look for a single B\'ezier curve that approximates multiple adjacent B\'ezier curves and satisfies certain conditions.
We propose the so-called \textit{box constraints}, which appear for the first time in the context of the merging problem.

A conventional problem of merging is to approximate multiple adjacent B\'ezier curves with a single B\'ezier curve which minimizes a selected
error function and satisfies some continuity constraints at the endpoints. Most of the papers deal with merging of only two B\'ezier curves (see \cite{HTJS01,Lu14,Lu13,THH03,ZW09}). Obviously, to merge more than two curves, one could use those algorithms repeatedly. However, such an approach increases the error of the approximation as well as the computational cost (see~\cite[\S1]{Lu15}). There are three methods that specialize in merging of more than two B\'ezier curves at the same time (see~\cite{CW08,Lu15,WGL}). Regardless of how many curves are merged, the most frequently used
strategy is to solve a system of normal equations (see,~e.g.,~\cite{Lu15}). In~\cite{WGL}, one can observe a different approach which is based on the properties of the so-called constrained dual Bernstein basis polynomials (to our knowledge, this method is the fastest one available).
The parametric (see,~e.g.,~\cite{CW08,Lu15,WGL}) or geometric (see,~e.g.,~\cite{Lu14,Lu15,ZW09}) continuity at the endpoints is preserved.

In~\cite{Gos15}, one of us proposed a new approach to the problem of degree reduction of B\'ezier curves.
The author noticed that as a result of the conventional degree reduction, the computed control points can be located far away from the plot of the curve.
He also explained why this is a serious defect. Next, to eliminate this issue, he solved the degree reduction problem with constraints of a new type.
In this paper, we show that the same observations may apply to the control points of the merged curve.
Therefore, the main goal of this paper is to formulate a new problem of merging of B\'ezier curves.
As in~\cite{Gos15}, the new approach requires completely different methods than in the case of the conventional one.

The outline of the paper is as follows. Further on in this section, we give necessary definitions and notation.
In Section~\ref{Sec:Pre}, we formulate the problem of merging of B\'{e}zier curves with box constraints.
The example motivating the restricted area approach is given in Section~\ref{Sec:motiv}.
Section~\ref{Sec:Mer} brings a solution of the new problem. Some illustrative examples are presented in Section~\ref{Sec:Ex}.
For a brief summary of the paper, see Section~\ref{Sec:Conc}.

Let $\Pi_m^d$ denote the space of all parametric polynomials in $\mathbb{R}^d$ of degree at most $m$; $\Pi_m:=\Pi^1_m$.
Further on in the paper, we use
$\mathbf{b}_{m,t} := \left[B_0^{m}(t),B_1^{m}(t),\ldots,B_m^{m}(t)\right]$, where
\begin{equation*}
    B_i^{m}(t) := \binom{m}{i}t^i(1-t)^{m-i} \qquad (i = 0, 1,\ldots, m;\ m \in \mathbb{N}),
\end{equation*}
to denote the vector of \textit{Bernstein polynomial basis} in $\Pi_m$.

We recall the well-known \textit{Gramian matrix} $\mathbf{G}_{m,n} := \left[g_{ij}\right] \in \mathbb{R}^{(m+1) \times (n+1)}$ of the Bernstein basis with the elements given by
$$
g_{ij} := \frac{1}{m+n+1}\binom{m}{i}\binom{n}{j}\rbinom{m+n}{i+j} \qquad (i=0,1,\ldots,m;\ j=0,1,\ldots,n).
$$

\textit{Forward difference operator} is defined by
$$
    \Delta^0q_i := q_i,\quad
    \Delta^jq_i := \Delta^{j-1}q_{i+1} - \Delta^{j-1}q_{i} \quad(j =1,2,\ldots).
$$

Let $\mathbf{M} \in \mathbb{R}^{n \times m}$ be a matrix, and let
$\mathcal{A} := \left\{i_1, i_2,\ldots, i_\alpha\right\} \subset \left[0,\,n-1\right]$, $\mathcal{B} := \left\{j_1, j_2,\ldots, j_\beta\right\} \subset \left[0,\,m-1\right]$ be the sets of natural numbers sorted in ascending order.
Notation
\begin{equation}\label{notMat}
\mathbf{M}^{\mathcal{A}, \mathcal{B}}
\end{equation}
defines a matrix formed by rows $i_1+1, i_2+1,\ldots, i_\alpha+1$ and columns $j_1+1, j_2+1,\ldots, j_\beta+1$ of the matrix $\mathbf{M}$.
Similarly, we use $\mathbf{v}^{\mathcal{A}}$, where $\mathbf{v}$ is a vector in $\mathbb{R}^{n}$.


\section{Problem of merging of B\'{e}zier curves with box constraints}\label{Sec:Pre}

In this section, we formulate the following new problem of merging of B\'{e}zier curves.

\begin{prob} \,[\textsf{Merging of B\'ezier curves with box constraints}]\label{P:Problem}\\
Let $0=t_0<t_1<\ldots<t_s=1$ be a partition of the interval $[0,\,1]$. Let there be given a \emph{composite B\'ezier curve} $P(t)$
($t\in[0,\,1]$) in $\R^d$, which in the interval $[t_{i-1},\,t_i]$ ($i=1,2,\ldots,s$) is exactly represented as a \emph{B\'ezier curve}
$P^i(t) \in \Pi_{n_i}^d$,
\[
 P(t)=P^i(t):=\sum_{j=0}^{n_i}p^i_j\,B^{n_i}_j(u_i(t)) \equiv \mathbf{b}_{n_i,u_i(t)}\mathbf{p}^i \qquad (t_{i-1}\le t\le t_i),
\]
where $u_i(t) := \frac{t-t_{i-1}}{\Delta t_{i-1}}$, and $\mathbf{p}^i := \left[p^i_0, p^i_1,\ldots, p^i_{n_i}\right]^T$ with $p^i_j := \left(p^{i,1}_j,p^{i,2}_j,\ldots,p^{i,d}_j\right) \in \mathbb{R}^d$.\\
\noindent Find a B\'ezier curve $R(t) \in \Pi_{m}^d$,
\[
R(t):=\sum_{j=0}^m r_j\,B^m_j(t) \equiv \mathbf{b}_{m,t}\mathbf{r} \qquad (0\le t\le 1),
\]
where $\mathbf{r} := \left[r_0, r_1,\ldots, r_m\right]^T$ with $r_j := \left(r_j^1,r_j^2,\ldots,r_j^d\right) \in \mathbb{R}^d$,
satisfying the following conditions:
\begin{itemize}
    \item[(i)] value of the squared \emph{$L_2$-error}
    \begin{equation}\label{min}
         E(\mathbf{r}) \equiv E^2_2 := \int_{0}^{1}\|P(t)-R(t)\|^2\dt,
    \end{equation}
    where $\|\cdot\|$ denotes the \emph{Euclidean vector norm} in $\mathbb{R}^d$,
    is minimized in the space $\Pi_m^d$;

    \item[(ii)] \emph{parametric continuity constraints at the endpoints} are satisfied, i.e.,
        \begin{equation}\label{cont}
	     \begin{array}{l}
		R^{(i)}(0)=P^{(i)}(0) \qquad (i=0,1,\ldots,k-1),\\[0.5ex]
		R^{(j)}(1)=P^{(j)}(1) \qquad (j=0,1,\ldots,l-1),
	     \end{array}
        \end{equation}
        where $k \leq n_1+1$, $l \leq n_s+1$, and $k+l\leq m$;

    \item[(iii)] control points $r_j$ $(k \leq j \leq m-l)$ are located inside the specified $d$-dimensional cube including the edges, i.e.,
    the following box constraints are fulfilled:
     \begin{equation}\label{box}
         c_h \leq r_{j}^h \leq C_h \qquad (j = k, k+1,\ldots, m-l;\ h= 1, 2,\ldots,d),
     \end{equation}
     where $c_h, C_h \in\mathbb R$.
\end{itemize}
\end{prob}

Notice that in the case of degree reduction of B\'ezier curves, analogical problem was formulated (cf.~\cite[Problem~3.1]{Gos15}).

\begin{remark}\label{R:Trad}
Note that papers \cite{Lu15, WGL} deal with the minimization of~\eqref{min}, with the conditions~\eqref{cont}, but without the box constraints~\eqref{box}.
In addition, a reasonable assumption that $m \ge \max_i{n_i}$ is made. Further on in this paper, such an approach is called the \textit{traditional merging}
(cf.~\cite[Remark~3.2]{Gos15}).
\end{remark}

\section{Motivation of the paper}\label{Sec:motiv}

As it turns out, the observations on the degree reduction problem (see~\cite[\S2]{Gos15}) also apply to the merging problem.
In order to see the issue clearly, let us consider the following example.

\begin{example}\label{Ex:1}
We give the planar composite B\'{e}zier curve ``Ampersand'', with three fifth degree B\'{e}zier segments (see Figure~\ref{figure1a}), defined by the control points $\{(0.49, 0.07),$ $(0.43, 0.22),$ $(0.08, 0.67),$ $(0, 0.97),$ $(0.29, 0.98),$ $(0.36, 0.9)\}$,
$\{(0.36, 0.9),$ $(0.43, 0.84),$ $(0.43, 0.68),$ $(0.25, 0.58),$ $(0.1, 0.36),$ $(0.09, 0.23)\}$, and
$\{(0.09, 0.23),$ $(0.08, 0.13),$ $(0.14, 0.06),$ $(0.34, 0),$ $(0.52, 0.08),$ $(0.48,$ $0.23)\}$, respectively. Assuming that the partition of the interval $[0,\,1]$ is given by $t_0 = 0,\ t_1 \doteq 0.45,\ t_2 \doteq 0.76,\ t_3 = 1$; we look for a single B\'{e}zier curve being the result of the traditional merging for $m = 14$, $k = 3$, $l = 1$ (see Remark~\ref{R:Trad}).

Figure~\ref{figure1b} shows the original composite curve and the merged curve. Clearly, the result of the approximation is very accurate.
The errors are $E_2 = 5.49e{-}03$ and
$E_{\infty} = 2.28e{-}02$, where
$$
E_{\infty} := \max_{t \in D_M} \|P(t) - R(t)\| \approx \max_{0 \leq t \leq 1} \|P(t) - R(t)\|
$$
with $D_M := \left\{0, 1/M, 2/M,\ldots, 1\right\}$ for $M := 500$. Observe also that the original control points are quite close to the plot of the curve (see Figure~\ref{figure1a}). In contrast, the resulting control points are located far away from the plot of the curve (see Figure~\ref{figure1c}). Note that we are unable to see the curve and its control points in one figure. Because of the non-intuitive location of the control points, further modeling of the merged curve is hard to imagine. A~designer that modifies the control points uses a convex hull property, which gives an intuition on shape and location of the curve. As it was stated in~\cite[\S2]{Gos15}, the size of the convex hull is a~measure of \textit{predictability} of the curve. Furthermore, let us recall that a small convex hull can be helpful while checking that two curves do not intersect, a curve and a surface do not intersect, a point does not lie on a curve. Observe that the convex hull of the resulting curve is huge, therefore, completely useless. Comparing this result with the ones from~\cite{Gos15}, we see that the defect seems to be even more significant (cf.~\cite[Figures 1b, 4b and 5b]{Gos15}).

Now, let us impose some box constraints. We want the searched control points to be inside the specified rectangular area (including edges of the rectangle).  Figure~\ref{figure2} presents the solution of Problem~\ref{P:Problem} for $m = 14$, $k = 3$, $l = 1$, $c_1 = -0.17$, $c_2 = 0$, $C_1 = 0.73$, $C_2 = 1.15$. Notice that the approximation is quite accurate (errors: $E_2 = 1.85e{-}02$ and $E_{\infty} = 6.10e{-}02$). Moreover, in this case, the computed control points are located much closer to the merged curve. As a result, the curve can be easily and intuitively modified by moving these points. What is more, we have obtained much smaller convex hull, which can be used to solve efficiently some important problems. More examples can be found in Section~\ref{Sec:Ex}.

\begin{figure}[H]
\captionsetup{margin=0pt, font={scriptsize}}
\begin{center}
\setlength{\tabcolsep}{0mm}
\begin{tabular}{c}
\subfloat[]{\label{figure1a}\includegraphics[width=0.32\textwidth]{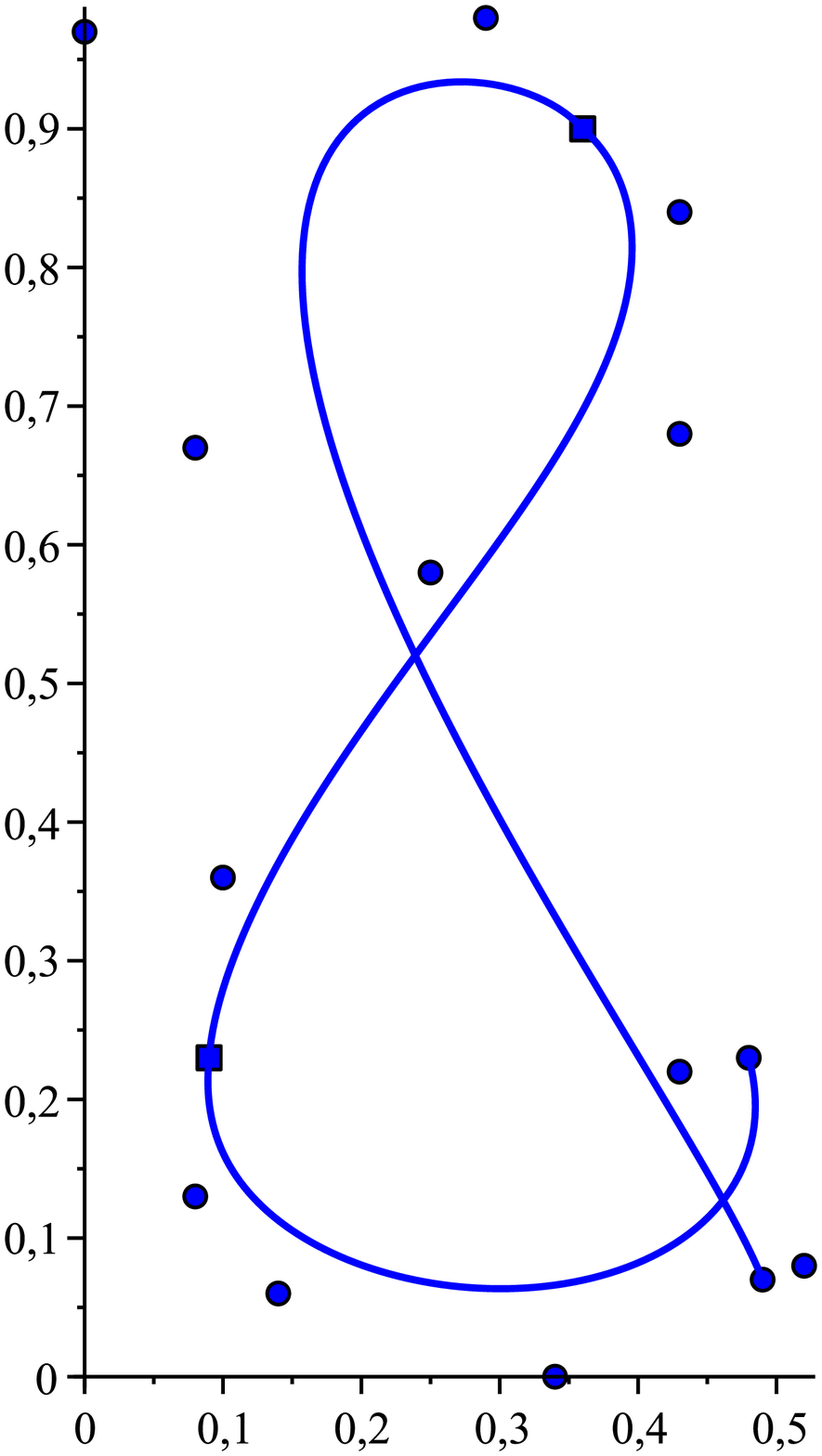}}
\subfloat[]{\label{figure1b}\includegraphics[width=0.302\textwidth]{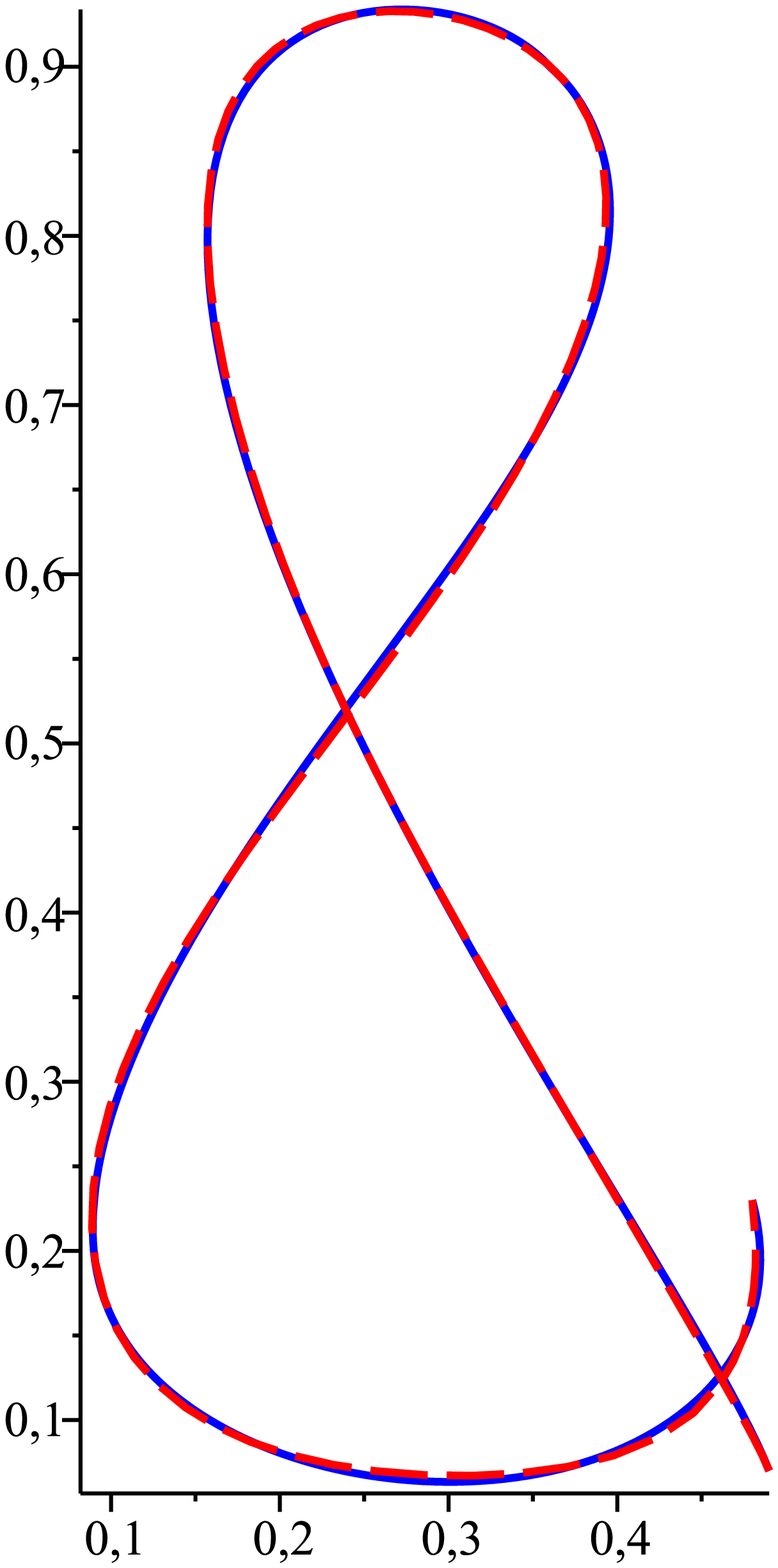}}
\subfloat[]{\label{figure1c}\includegraphics[width=0.281\textwidth]{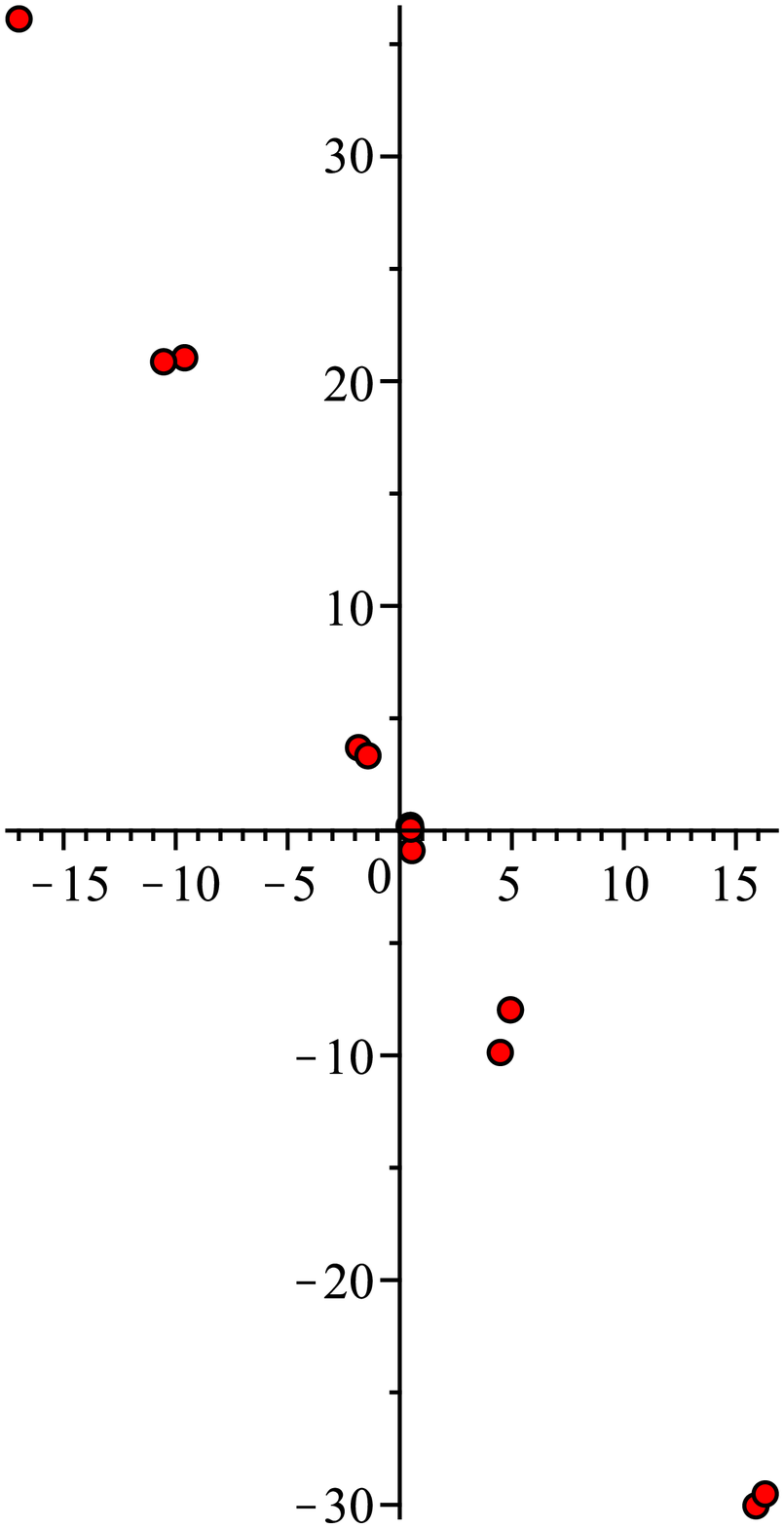}}
\end{tabular}
    \caption{Figure~(a) shows the original composite B\'ezier curve with its control points. Figure~(b) illustrates the original composite B\'ezier curve (blue solid line), and the merged B\'ezier curve (red dashed line) being the solution of the traditional merging problem. Figure~(c) presents the control points of the merged curve (red color).}
\end{center}
\end{figure}
\end{example}

\begin{figure}[H]
   \captionsetup{font=scriptsize}
   \centering
   \includegraphics[width=73mm]{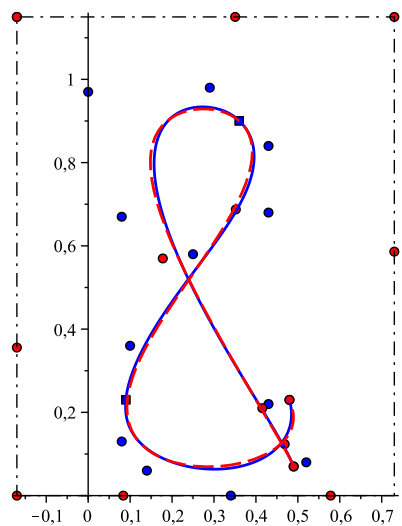}
   \caption{The original composite B\'ezier curve (blue solid line with blue control points), and the merged B\'ezier curve (red dashed line with red control points) being the solution of Problem~\ref{P:Problem}. See the restricted area (black dotted-dashed frame).}
   \label{figure2}
\end{figure}

\section{Merging of B\'{e}zier curves with box constraints}\label{Sec:Mer}

Now, we give the method of solving Problem~\ref{P:Problem}.

First, we notice that some observations concerning the box-constrained degree reduction are also true in the case of the box-constrained merging.
Clearly, a B\'ezier curve being the solution of Problem~\ref{P:Problem} can be obtained in a componentwise
way (cf.~\cite[Remark~3.3]{Gos15}). 
Therefore, it is sufficient to describe our method in the case of $d = 1$. Further on in this section, we assume that $P^i \in \Pi_{n_i}$ $(i=1,2,\ldots,s)$,
$R \in \Pi_m$, and $c, C \in \R$ are the lower and upper bounds for the box constraints~\eqref{box}.

Next, we recall that the conditions~\eqref{cont} yield the following well-known formulas (see,~e.g.,~\cite[Theorem~3.1]{WGL}):
\begin{align*}
	  & \displaystyle
	  r_j=\binom{n_1}{j}\binom{m}{j}^{\!-1}\,\Delta^jp^{1}_0-
          \sum_{h=0}^{j-1}(-1)^{j+h}\binom{j}{ h}r_{h}\qquad (j=0,1,\ldots,k-1),\\[1ex]
	& \displaystyle r_{m-j}
	=(-1)^j\binom{n_s}{j}\binom{m}{j}^{\!-1}\,\Delta^jp^{s}_{n_s-j}-
	\sum_{h=1}^{j}(-1)^h\binom{j}{ h}r_{m-j+h}\qquad (j=0,1,\ldots,l-1).		  	
\end{align*}

What remains is to minimize $E(\mathbf{r})$ subject to the conditions~\eqref{box} for $d=1$.
One can see clearly that $E(\mathbf{r})$ is a quadratic function. Therefore, in the following subsection,
we are dealing with the so-called \textit{box-constrained quadratic programming problem}.

\subsection{Quadratic programming with box constraints}\label{SubSubSec:QP}

In this subsection, we use the quadratic programming approach to solve Problem~\ref{P:Problem}.

Quadratic programming is an \textit{optimization problem} of minimizing or maximizing a quadratic \textit{objective function} of several variables subject to linear constraints on these variables. Taking into account the particular form of the restrictions~\eqref{box}, let us consider the following quadratic programming problem with box constraints:
\begin{equation}\label{E:QP}
\min_{}\;{\frac{1}{2}\mathbf{x}^T\mathbf{Q}\mathbf{x} + \mathbf{x}^T\mathbf{d}}, \qquad \mbox{s.t.} \qquad c \leq x_j \leq C,
\end{equation}
where $\mathbf{d}, \mathbf{x}:=\left[x_j\right] \in \mathbb{R}^{m-k-l+1}$ and $\mathbf{Q} \in \mathbb{R}^{(m-k-l+1) \times (m-k-l+1)}$.
In our case, we set $\mathbf{x} := \mathbf{r}^{\mathcal{F}}$, where we define $\mathcal{F} := \left\{k, k+1,\ldots,m-l\right\}$ and use the notation of~\eqref{notMat}. Now, we will adjust $E(\mathbf{r})$ to the form~\eqref{E:QP}.

First, taking into account that $P$ is a piecewise polynomial, we have to subdivide the searched polynomial $R$ as well. This can be done by applying the de Casteljau algorithm. In~\cite[\S2]{Lu15}, Lu gave the following formula:
$$
R(t) = R^i(t) := \mathbf{b}_{m,u_i(t)}\mathbf{D}_{i}\mathbf{r} \qquad (t_{i-1}\le t\le t_i;\ i=1,2,\ldots,s),
$$
where
\begin{equation}\label{Eq:Di}
\mathbf{D}_i := \mathbf{A}_1(t_{i-1}/t_i)\mathbf{A}_2(t_i)
\end{equation}
with
$$
 \mathbf{A}_1(\lambda) =
 \begin{bmatrix}
  B_0^m(\lambda) & B_1^m(\lambda) & \cdots & B_m^m(\lambda) \\
  0 & B_0^{m-1}(\lambda) & \cdots & B_{m-1}^{m-1}(\lambda) \\
  \vdots  & \vdots  & \ddots & \vdots  \\
  0 & 0 & \cdots & 1
 \end{bmatrix},\quad
\mathbf{A}_2(\lambda) =
 \begin{bmatrix}
  1 & 0 & \cdots & 0 \\
  B_0^1(\lambda) & B_1^1(\lambda) & \cdots & 0 \\
  \vdots  & \vdots  & \ddots & \vdots  \\
  B_0^m(\lambda) & B_1^m(\lambda) & \cdots & B_m^m(\lambda)
 \end{bmatrix}.
$$

\begin{remark}\label{R:d}
According to~\cite[Lemmas~2.5,\;2.4]{WGL}, $\mathbf{D}_i = \left[d^{(i)}_{jh}\right] \in \mathbb{R}^{(m+1)\times(m+1)}$,
where the entries $d^{(i)}_{jh}$ $(i = 1, 2,\ldots, s;\ j = 0, 1,\ldots, m;\ h = 0, 1,\ldots, m)$
satisfy the following recurrence relation:
\begin{multline*}
 \Delta t_{i-1}\left[(m-j+1)d_{j-1,h}^{(i)}+(2j-m)d_{jh}^{(i)}-(j+1)d_{j+1,h}^{(i)}\right]\\
\qquad\quad = (m-h)d_{j,h+1}^{(i)} + (2h-m)d_{jh}^{(i)} - hd_{j,h-1}^{(i)}\\
 (1 \leq j \leq m-1;\ 0 \leq h \leq m).
\end{multline*}
Therefore, one can avoid matrix multiplications and compute the matrices $\mathbf{D}_1,\mathbf{D}_2,\ldots,\mathbf{D}_s$ efficiently, with the complexity $O(sm^2)$, using~\cite[Algorithm~4.1]{WGL}. Observe that the direct use of~\eqref{Eq:Di} results in the complexity $O(sm^3)$.
\end{remark}

Next, assuming that $\mathcal{K} := \left\{0,1,\ldots,m\right\}$, $\mathcal{C} := \mathcal{K}\setminus \mathcal{F}$ and using the notation of~\eqref{notMat}, we write (cf.~\cite[(11)]{Lu15})
\begin{align*}
E(\mathbf{r}) &= \int_{0}^{1}(P(t)-R(t))^2\dt = \sum_{i=1}^s\int_{t_{i-1}}^{t_i}\left(P^i(t)-R^i(t)\right)^2\dt \\
&= \sum_{i=1}^s\Delta t_{i-1} \int_{0}^{1} \left(\mathbf{b}_{n_i,v}\mathbf{p}^{i} - \mathbf{b}_{m,v}\mathbf{D}_i^{\mathcal{K},\mathcal{C}}\mathbf{r}^{\mathcal{C}}
- \mathbf{b}_{m,v}\mathbf{D}_i^{\mathcal{K},\mathcal{F}}\mathbf{r}^{\mathcal{F}}\right)^2\dv \\
&= \frac{1}{2}\left(\mathbf{r}^{\mathcal{F}}\right)^T\mathbf{Q}\mathbf{r}^{\mathcal{F}} +
  \left(\mathbf{r}^{\mathcal{F}}\right)^T\mathbf{d} + a \ =: \  g\left(\mathbf{r}^{\mathcal{F}}\right) + a,
\end{align*}
where
\begin{align*}
& \mathbf{Q} := 2\sum_{i=1}^s\Delta t_{i-1}\left(\mathbf{D}_i^{\mathcal{K},\mathcal{F}}\right)^T\mathbf{G}_{m,m}\mathbf{D}_i^{\mathcal{K},\mathcal{F}},\\
& \mathbf{d} :=  2\sum_{i=1}^s\Delta t_{i-1}\left(\mathbf{D}_i^{\mathcal{K},\mathcal{F}}\right)^T
\left(\mathbf{G}_{m,m}\mathbf{D}_i^{\mathcal{K},\mathcal{C}}\mathbf{r}^{\mathcal{C}} - \mathbf{G}_{m,n_i}\mathbf{p}^{i}\right),
\end{align*}
and $a \in \mathbb{R}$ is a certain constant term. Obviously, $a$ is meaningless in the minimization process, therefore, the \textit{significant terms} of $E(\mathbf{r})$ are given by $g(\mathbf{r}^{\mathcal{F}})$, which is written in the form~\eqref{E:QP}.

\begin{remark}\label{R:Sol}
Matrix $\mathbf{Q}$ is positive definite (see~\cite[\S3.1]{Lu15}), therefore, the objective function $g$ is strictly convex.
Furthermore, the feasible set is nonempty, closed and convex. We conclude that the quadratic programming problem has a~unique solution (see, e.g.,~\cite[Proposition~2.5]{DOS}) and so does Problem~\ref{P:Problem}. In contrast, a solution of the analogical degree reduction problem
may not be unique (cf.~\cite[Theorem~4.1]{Gos15}). The difference is that, in the present paper, we consider the \textit{continuous inner product}
(see~\eqref{min}) instead of the \textit{discrete inner product} (see~\cite[(3.1)]{Gos15}).
\end{remark}

There are many papers dealing with the box-constrained quadratic programming problem. To solve it, one can use a variety of strategies, including
\textit{active set methods} (see, e.g.,~\cite{Fer98}) and \textit{interior point algorithms} (see, e.g.,~\cite{Han90}). Some of the approaches combine the active set strategy with \textit{gradient projection method} (see, e.g.,~\cite{Mor89}). For extensive lists of references, see the mentioned papers.

\section{Examples}\label{Sec:Ex}

In this section, we apply our method to the composite B\'ezier curves in $\R^2$.

As in~\cite{WGL}, we generalize the approach of \cite{Lu14} and obtain a partition of the interval $[t_0,\,t_s]=[0,\,1]$ according to the lengths of segments $P^i$:
\begin{equation}\label{E:part}
t_j :=L_j/L_s  \qquad (j=1,2,\ldots,s-1),
\end{equation}
where
\[
L_q:=\sum_{i=1}^{q}\int_{0}^{1}\left\Vert\frac{\der}{\dt}\sum_{h=0}^{n_i}p^i_hB^{n_i}_h(t)\right\Vert\dt.
\]
Integrals are evaluated using $\mbox{Maple}{\small \texttrademark}$ \texttt{int} procedure with the option \texttt{numeric}.

A solution of the traditional merging problem (see Remark~\ref{R:Trad}) is computed using~\cite[Algorithm~4.2]{WGL}.
The complexity of this algorithm is $O(sm^2)$ which, to our knowledge, is significantly less than cost of other methods of merging with the constraints~\eqref{cont} (cf.~\cite{CW08,Lu15}).
To solve the quadratic programming problem with box constraints~\eqref{E:QP}, we use the matrix version of Maple{\small \texttrademark} \texttt{QPSolve} command. It is worth noting that this procedure implements an iterative active set method and it is suited for the box constraints, i.e., the vectors of lower and upper bounds can be passed using the optional parameter \texttt{bd}. According to the documentation provided by $\mbox{Maplesoft}{\small \texttrademark}$, in the case of the convex optimization, a global minimum is returned (cf. Remark~\ref{R:Sol}). For the initial point, we choose the lower bounds, i.e., $c_1$ and $c_2$.

The results have been obtained on a computer with \texttt{Intel Core i5-3337U 1.8GHz} processor and \texttt{8GB} of \texttt{RAM}, using $24$-digit arithmetic. $\mbox{Maple}{\small \texttrademark}13$ worksheet containing programs and tests is available at \url{http://www.ii.uni.wroc.pl/~pgo/papers.html}.

\begin{example}\label{Ex:2}
We introduce the composite B\'ezier curve ``D'' (see Figure~\ref{figure3a}), formed by three cubic segments which are defined by the control points
$\{(0.32, 0.81),$ $(0.26, 0.59),$ $(0.18, 0),$ $(0.06, 0.27)\}$,
$\{(0.06, 0.27),$ $(0, 0.42),$ $(0.42, 0.08),$ $(0.57, 0.25)\}$ and
$\{(0.57, 0.25),$ $(0.76, 0.46),$ $(0.8,$ $1),$ $(0.22, 0.85)\}$, respectively.
Formula \eqref{E:part} implies $t_0 = 0,\ t_1 \doteq 0.32,\ t_2 \doteq 0.57,\ t_3 = 1$.
Figure~\ref{figure3b} shows the result of the traditional merging for $m = 18$, $k = 1$, $l = 2$.
The merged curve looks like a perfect approximation (errors: $E_2 = 3.35e{-}03$ and $E_{\infty}= 9.57e{-}03$), unfortunately, it suffers from the defect described in Section~\ref{Sec:motiv} (see Figure~\ref{figure3c}). To avoid this, we solve Problem~\ref{P:Problem} for $m = 18$, $k = 1$, $l = 2$, with the following box constraints:
\begin{equation}\label{Eq:Res1}
\begin{aligned}
    &c_1 := \min_{1 \leq i \leq s}\min_{0 \leq j \leq n_i}p_j^{i,1} - 0.2 = -0.2,\qquad
    &&C_1 := \max_{1 \leq i \leq s}\max_{0 \leq j \leq n_i}p_j^{i,1} = 0.8,\\
    &c_2 := \min_{1 \leq i \leq s}\min_{0 \leq j \leq n_i}p_j^{i,2} - 0.3 = -0.3,\qquad
    &&C_2 := \max_{1 \leq i \leq s}\max_{0 \leq j \leq n_i}p_j^{i,2} = 1
\end{aligned}
\end{equation}
(cf.~\eqref{box}), and obtain the curve shown in Figure~\ref{figure3d} (errors: $E_2 = 1.38e{-}02$ and $E_{\infty}= 2.98e{-}02$). Compare Figure~\ref{figure3d} with Figure~\ref{figure3c} to see a big difference in the location of the resulting control points. Obviously, the curve in Figure~\ref{figure3d} is much more satisfying in this regard.

\begin{figure}[H]
\captionsetup{margin=0pt, font={scriptsize}}
\begin{center}
\setlength{\tabcolsep}{0mm}
\begin{tabular}{c}
\subfloat[]{\label{figure3a}\includegraphics[width=0.346\textwidth]{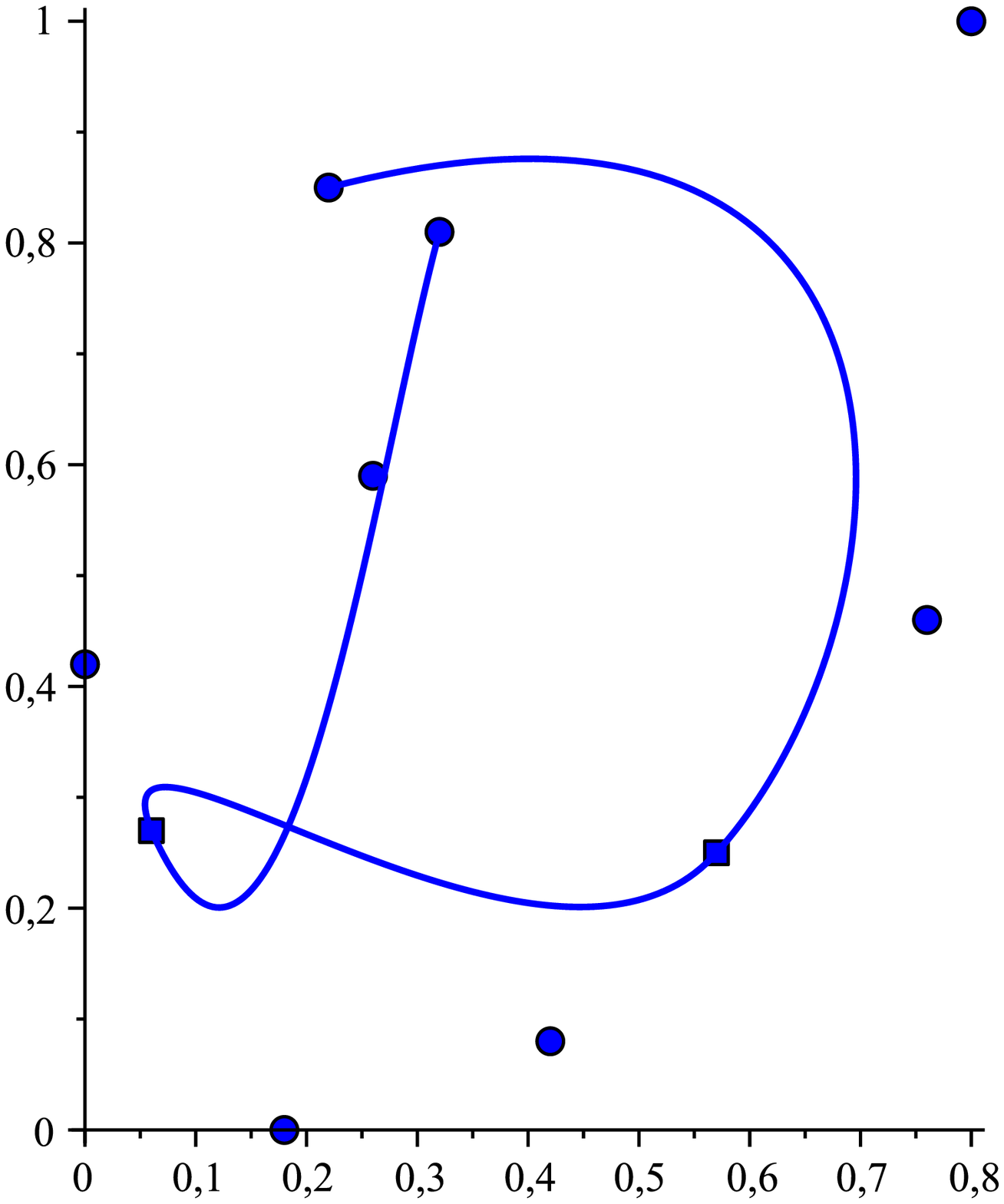}}
\subfloat[]{\label{figure3b}\includegraphics[width=0.388\textwidth]{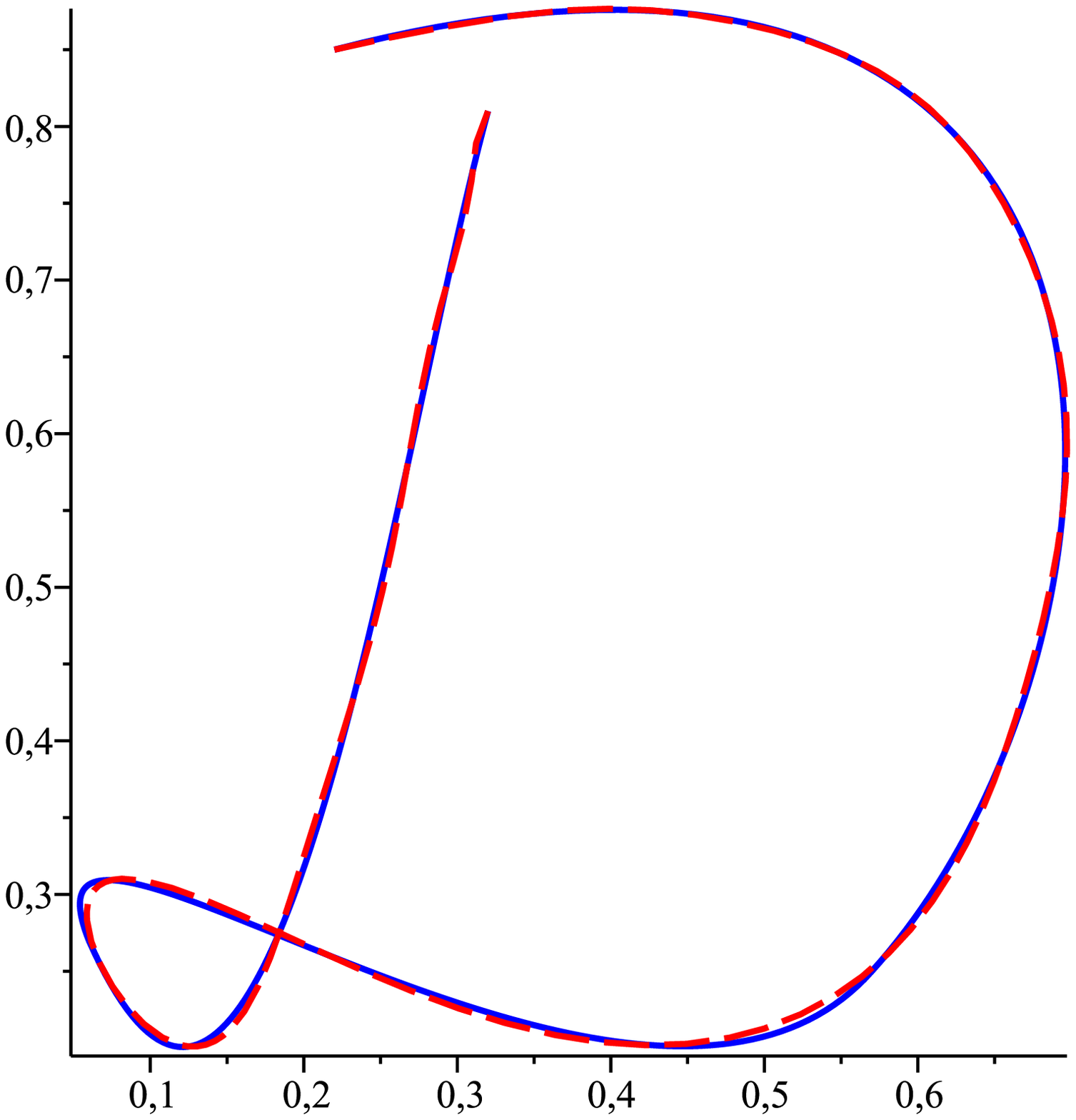}}\\
\subfloat[]{\label{figure3c}\includegraphics[width=0.65\textwidth]{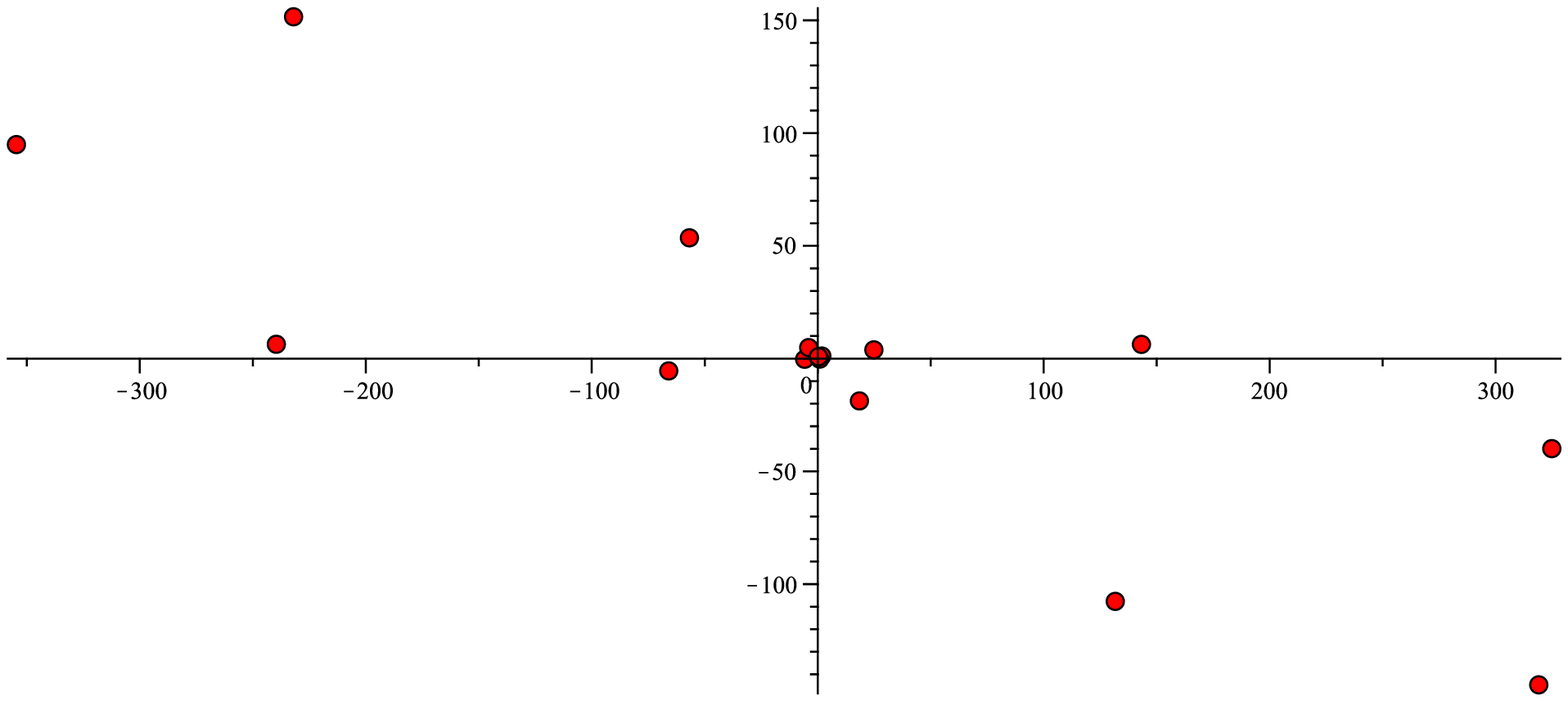}}\\
\subfloat[]{\label{figure3d}\includegraphics[width=0.346\textwidth]{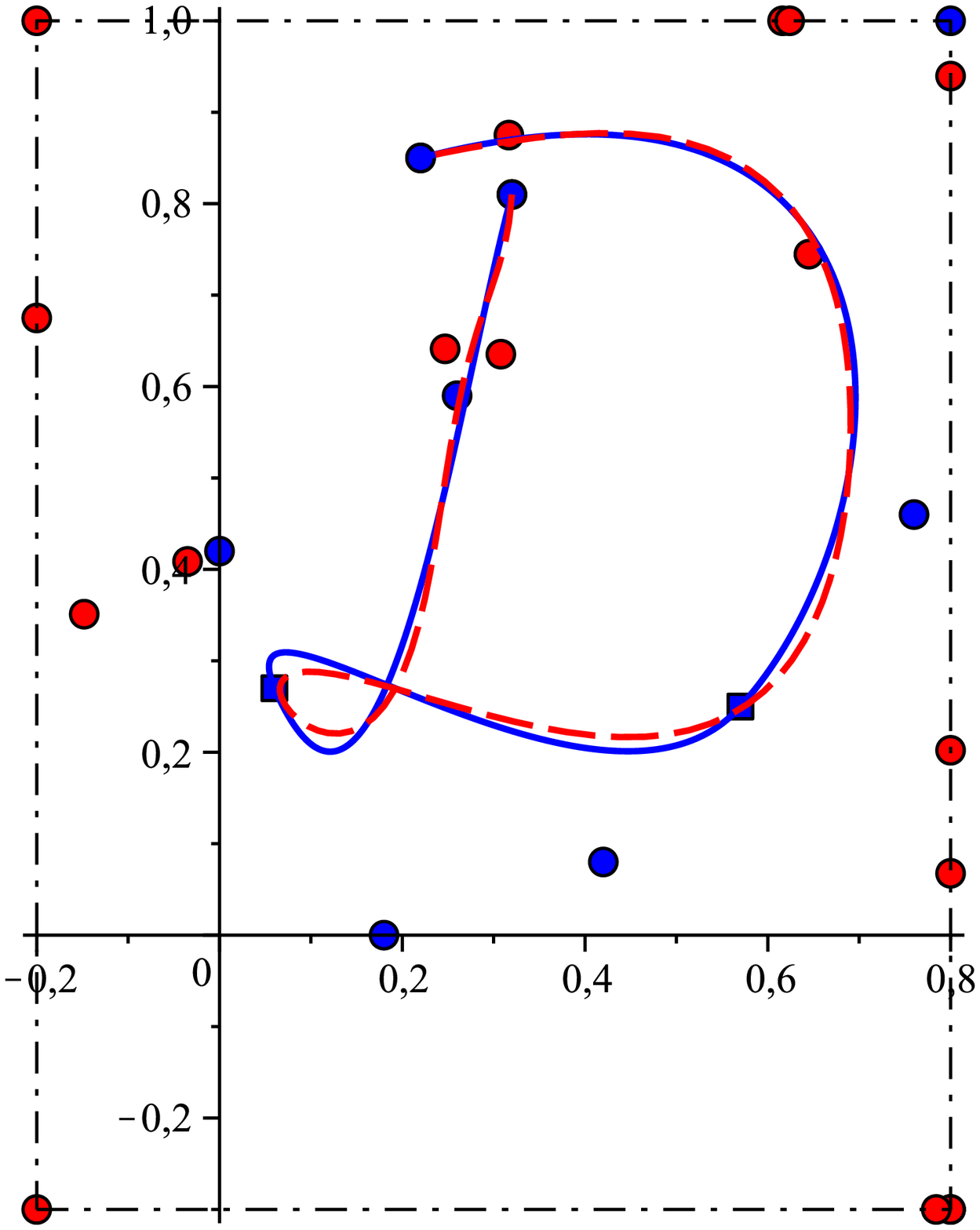}}
\end{tabular}
    \caption{Merging of three segments of the composite B\'{e}zier curve. The original composite curve (blue solid line with blue control points) and the merged curve (red dashed line with red control points), parameters: $m = 18$, $k = 1$, $l = 2$. Figure~(a) shows the original composite curve with its control points. Figure~(b) illustrates the curve being the solution of the traditional merging problem. Figure~(c) presents the control points of the merged curve shown in Figure~(b). The curve being the solution of~Problem~\ref{P:Problem} with the resulting control points and the restricted area (black dotted-dashed frame) are shown in Figure~(d).}
\end{center}
\end{figure}

\end{example}

\begin{figure}[H]
\captionsetup{margin=0pt, font={scriptsize}}
\begin{center}
\setlength{\tabcolsep}{0mm}
\begin{tabular}{c}
\subfloat[]{\label{figure4a}\includegraphics[width=0.386\textwidth]{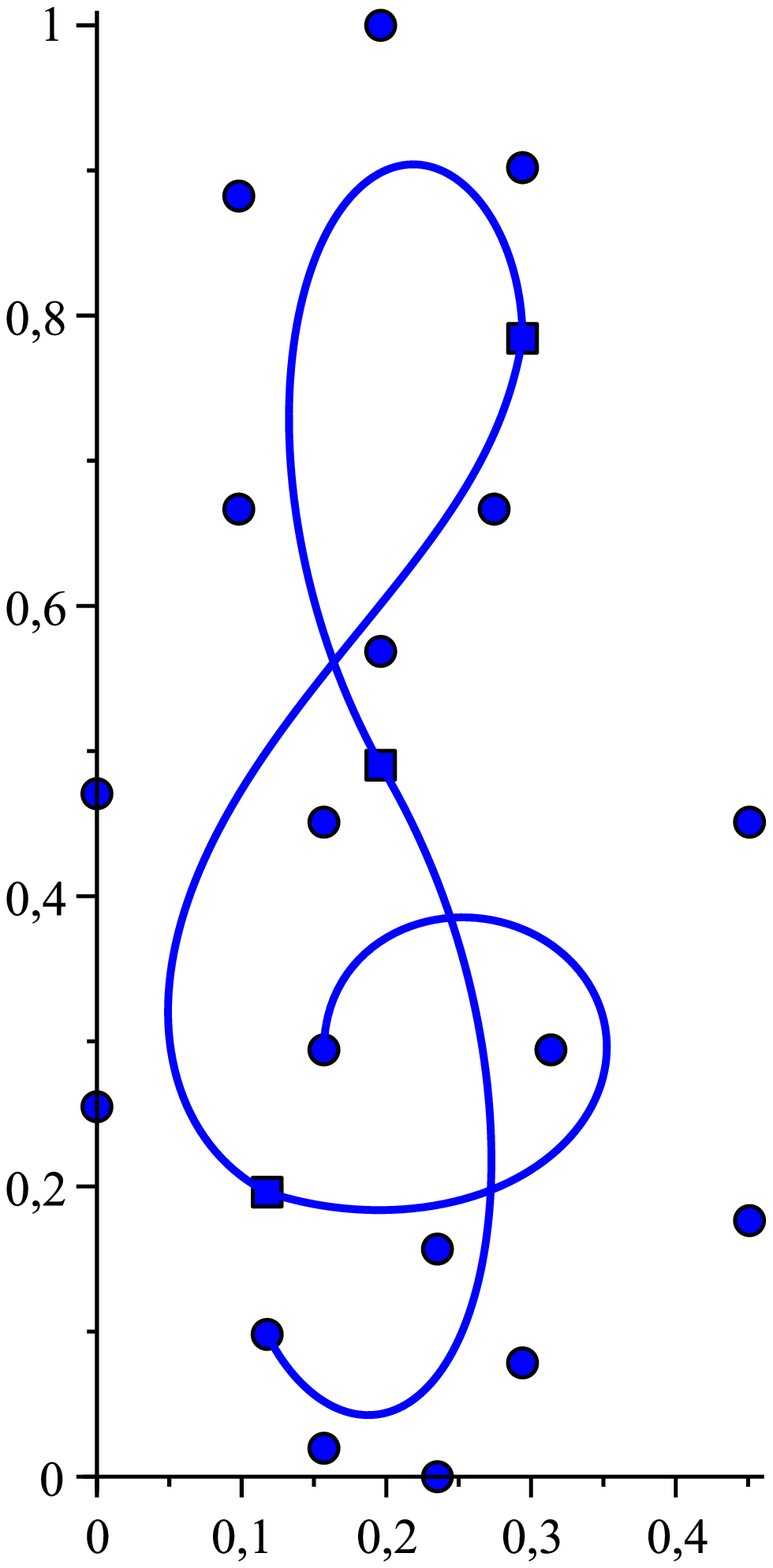}}
\subfloat[]{\label{figure4b}\includegraphics[width=0.388\textwidth]{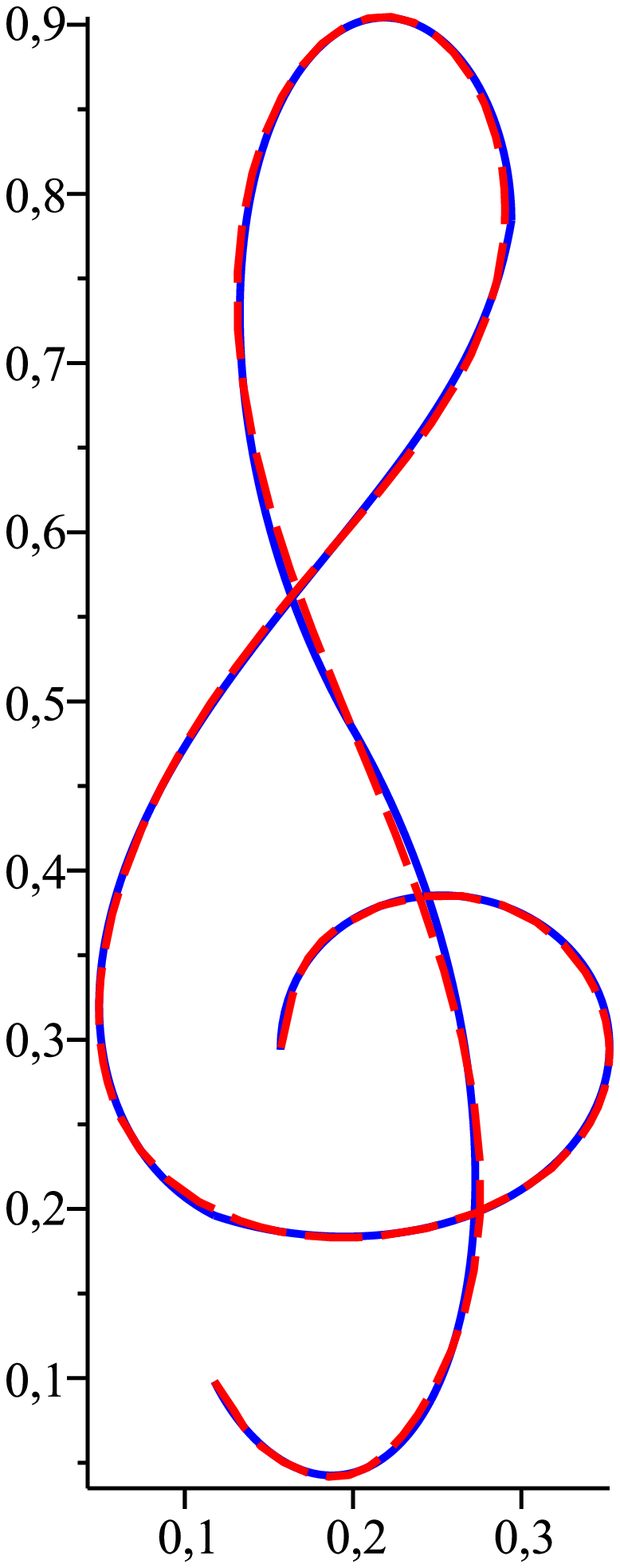}}\\
\subfloat[]{\label{figure4c}\includegraphics[width=0.439\textwidth]{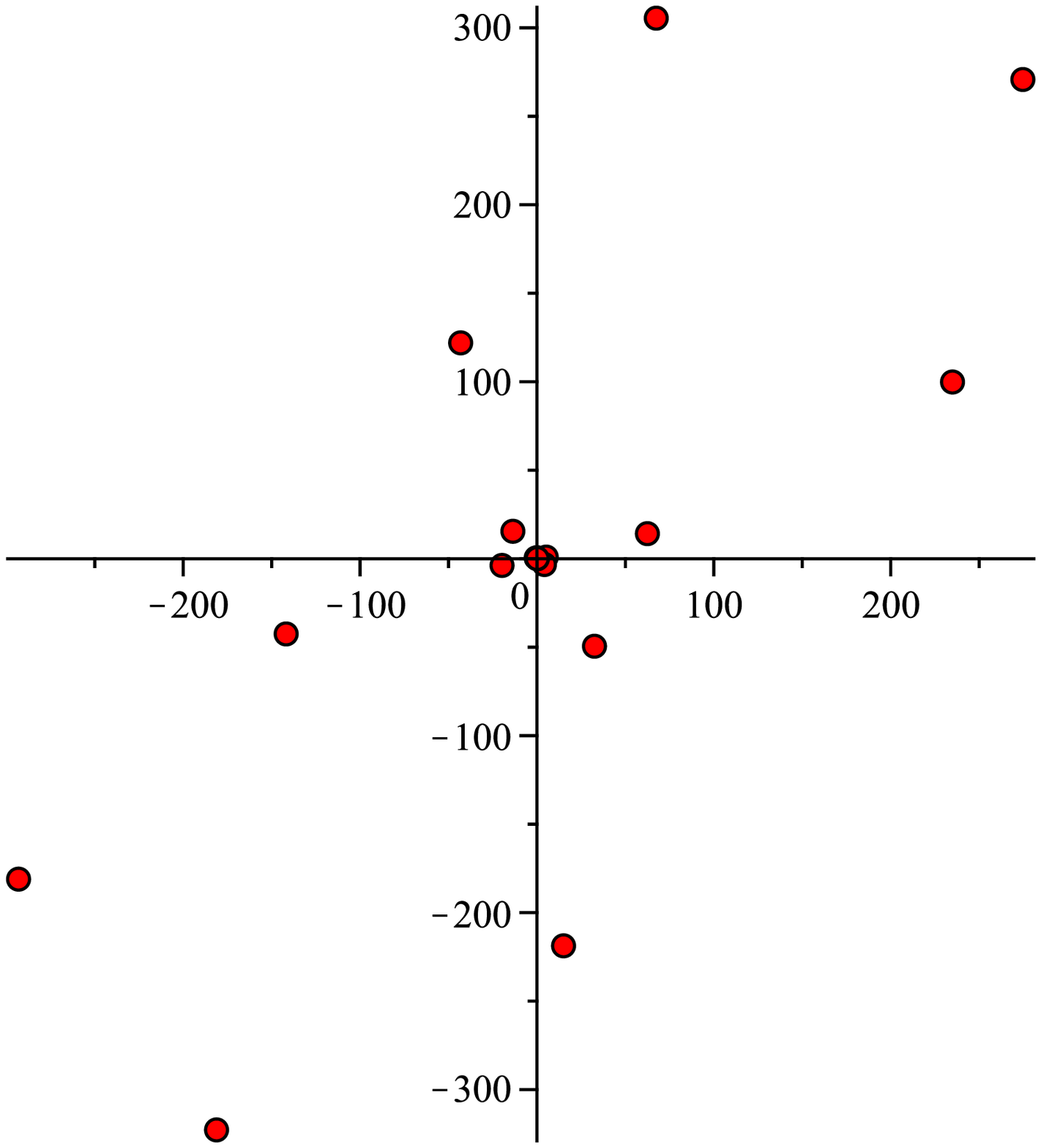}}
\subfloat[]{\label{figure4d}\includegraphics[width=0.388\textwidth]{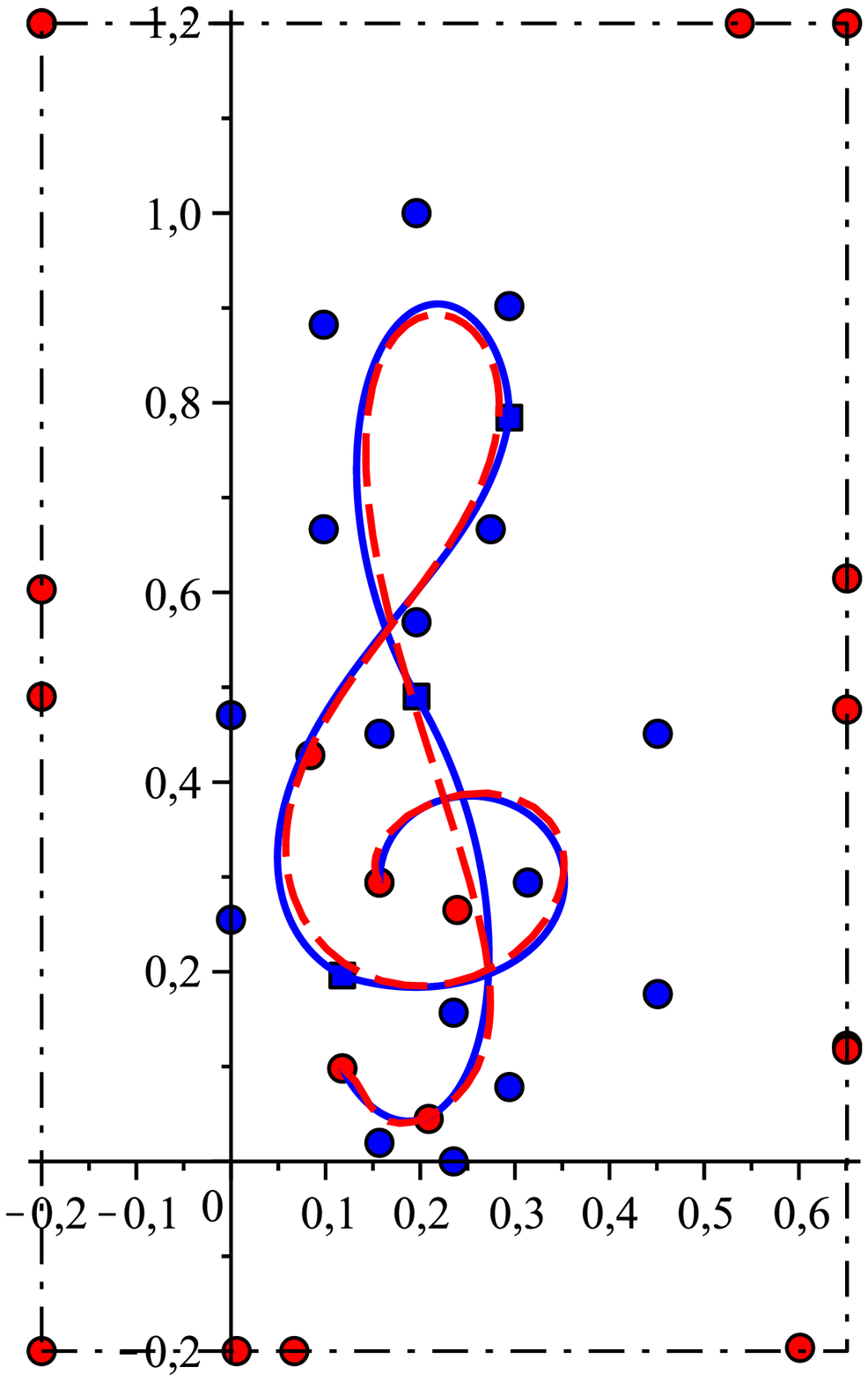}}
\end{tabular}
    \caption{Merging of four segments of the composite B\'{e}zier curve. The original composite curve (blue solid line with blue control points) and the merged curve (red dashed line with red control points), parameters: $m = 19$, $k = l = 1$. Figure~(a) shows the original composite curve with its control points. Figure~(b) illustrates the curve being the solution of the traditional merging problem. Figure~(c) presents the control points of the merged curve shown in Figure~(b). The curve being the solution of~Problem~\ref{P:Problem} with the resulting control points and the restricted area (black dotted-dashed frame) are shown in Figure~(d).}
\end{center}
\end{figure}

\begin{example}\label{Ex:3}
Now, we consider the composite B\'ezier curve with four fifth degree B\'{e}zier segments (see Figure~\ref{figure4a}).
For the original control points, see~\cite[Example 3]{Lu15}. To place the curve inside the unit box, we have divided each coordinate of the control points by $5.1$. According to~\eqref{E:part}, we get
$t_0 = 0,\ t_1 \doteq 0.24,\ t_2 \doteq 0.49,\ t_3 \doteq 0.76,\ t_4 = 1$.
As a result of the traditional merging ($m = 19$, $k = l = 1$), we obtain the B\'ezier curve which is illustrated in Figure~\ref{figure4b}.
Once again, we get a good approximation (errors: $E_2 = 2.08e{-}03$ and $E_{\infty}= 5.65e{-}03$), however, the resulting control points
are located far away from the plot of the curve (see Figure~\ref{figure4c}). Taking into account the axis scale in Figure~\ref{figure4c}, we conclude that this example seems to be extremely difficult. Nonetheless, the solution of Problem~\ref{P:Problem} for $m = 19$, $k = l = 1$, with the box constraints
 \begin{equation*}
\begin{aligned}
    &c_1 := \min_{1 \leq i \leq s}\min_{0 \leq j \leq n_i}p_j^{i,1} - 0.2 = -0.2,\qquad
    &&C_1 := \max_{1 \leq i \leq s}\max_{0 \leq j \leq n_i}p_j^{i,1} + 0.2 \doteq 0.65,\\
    &c_2 := \min_{1 \leq i \leq s}\min_{0 \leq j \leq n_i}p_j^{i,2} - 0.2 = -0.2,\qquad
    &&C_2 := \max_{1 \leq i \leq s}\max_{0 \leq j \leq n_i}p_j^{i,2} + 0.2 = 1.2
\end{aligned}
\end{equation*}
is quite decent (errors: $E_2 = 9.71e{-}03$ and $E_{\infty}= 1.90e{-}02$). See Figure~\ref{figure4d}.
\end{example}

\begin{remark}\label{R:Res}
As stated in~\cite[Remark~6.3]{Gos15}, selection of the restricted area is a difficult issue. The choice always depends on the considered
example and on the precision level that we accept as satisfactory. However, there is a strategy that seems to work quite well for the given examples.
To explain this procedure, let us revisit Example~\ref{Ex:2}. At the beginning, we set
\begin{equation}\label{Eq:DBox1}
\begin{aligned}
    &c^{(1)}_1 := \min_{1 \leq i \leq s}\min_{0 \leq j \leq n_i}p_j^{i,1}=0,\qquad
    &&C^{(1)}_1 := \max_{1 \leq i \leq s}\max_{0 \leq j \leq n_i}p_j^{i,1}=0.8,\\
    &c^{(1)}_2 := \min_{1 \leq i \leq s}\min_{0 \leq j \leq n_i}p_j^{i,2}=0,\qquad
    &&C^{(1)}_2 := \max_{1 \leq i \leq s}\max_{0 \leq j \leq n_i}p_j^{i,2}=1.
\end{aligned}
\end{equation}
Consequently, the resulting control points will be bounded by the outermost control points of the original curves.
Unfortunately, the obtained curve is unsatisfactory (see~Figure~\ref{figure5a}). Next, to improve this result, we must expand the
restricted area. Intuition tells us that we should try to move the borders with the highest numbers of the control points. We consider
\begin{equation}\label{Eq:DBox2}
\begin{aligned}
    c^{(2)}_1 := c^{(1)}_1 - 0.04w_1 \doteq -0.05,\qquad
    C^{(2)}_1 := C^{(1)}_1,\\
    c^{(2)}_2 := c^{(1)}_2 - 0.04w_1 \doteq -0.05,\qquad
    C^{(2)}_2 := C^{(1)}_2,
\end{aligned}
\end{equation}
where
$$
w_i := \sqrt{\left(C^{(i)}_1 - c^{(i)}_1\right)^2+\left(C^{(i)}_2 - c^{(i)}_2\right)^2}
$$
is the diagonal length of $i$-th restricted area. Notice that the error is now lower (see~Figure~\ref{figure5b} and Table~\ref{tab:table1}).
Therefore, we should try to make another step in the same direction. This time, the expansion is greater, i.e., we set
\begin{equation}\label{Eq:DBox3}
\begin{aligned}
    c^{(3)}_1 := c^{(2)}_1 - 0.08w_2 \doteq -0.16,\qquad
    C^{(3)}_1 := C^{(2)}_1,\\
    c^{(3)}_2 := c^{(2)}_2 - 0.08w_2 \doteq -0.16,\qquad
    C^{(3)}_2 := C^{(2)}_2.
\end{aligned}
\end{equation}
The result can be seen in Figure~\ref{figure5c}. See also Table~\ref{tab:table1}. Observe that, in Example~\ref{Ex:2}, the restricted area~\eqref{Eq:Res1} is even larger. 
See Figure~\ref{figure3d}. Taking into account that \texttt{QPSolve} is an iterative method which we apply separately for each coordinate, pairs of numbers of iterations are also given in Table~\ref{tab:table1}.

According to our experiments, if the control points of the optimal solution of the traditional merging are located very far away from the plot of the curve (see Figures~\ref{figure1c}, \ref{figure3c} and \ref{figure4c}), then it is difficult to find a satisfying solution of Problem~\ref{P:Problem}.
For that reason, the examples given in this paper are much more demanding than the ones presented in~\cite{Gos15}. Moreover, note that in the case of the box-constrained merging, majority of the resulting control points are located on borders (see Figures~\ref{figure2}, \ref{figure3d} and \ref{figure4d}).

Regardless of choice of the restricted area, one should realize that because of the additional constraints~\eqref{box}, approximation error must be inevitably larger than for the traditional approach.

\begin{table}[H]
\captionsetup{margin=0pt, font={scriptsize}}
\ra{1.3}
\centering
\scalebox{1}{
\begin{tabular}{@{}ccccccc@{}}
\toprule
Box constraints & \phantom{ab} & $E_2$ & \phantom{ab} & $E_{\infty}$  &\phantom{ab} & Iterations
\\ \midrule
\eqref{Eq:DBox1} &  & $2.25e{-}02$ & & $5.54e{-}02$ & & $(19,19)$\\
\eqref{Eq:DBox2} &  & $1.86e{-}02$ & & $4.14e{-}02$ & & $(19,19)$\\
\eqref{Eq:DBox3} &  & $1.51e{-}02$ & & $3.30e{-}02$ & & $(17,18)$\\
\eqref{Eq:Res1} &  & $1.38e{-}02$ & & $2.98e{-}02$ & & $(22,29)$\\
\hline
\end{tabular}}
\caption{$L_2$-errors, maximum errors and numbers of iterations for merging of three segments of the composite B\'{e}zier curve ``D'' with box constraints.
Parameters: $m = 18$, $k = 1$, $l = 2$.}
\label{tab:table1}
\end{table}

\begin{figure}[H]
\captionsetup{margin=0pt, font={scriptsize}}
\begin{center}
\setlength{\tabcolsep}{0mm}
\begin{tabular}{c}
\subfloat[]{\label{figure5a}\includegraphics[width=0.335\textwidth]{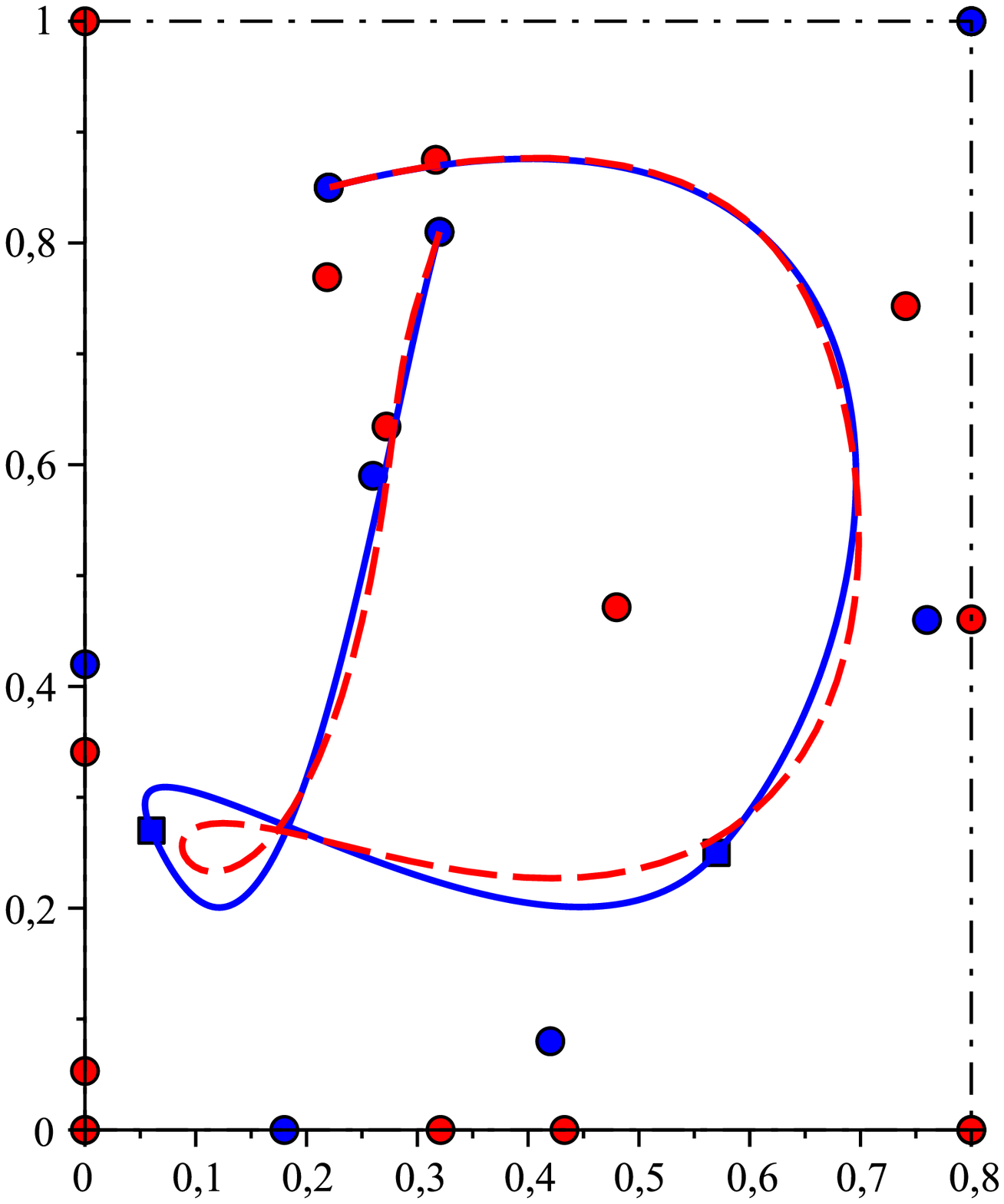}}
\subfloat[]{\label{figure5b}\includegraphics[width=0.335\textwidth]{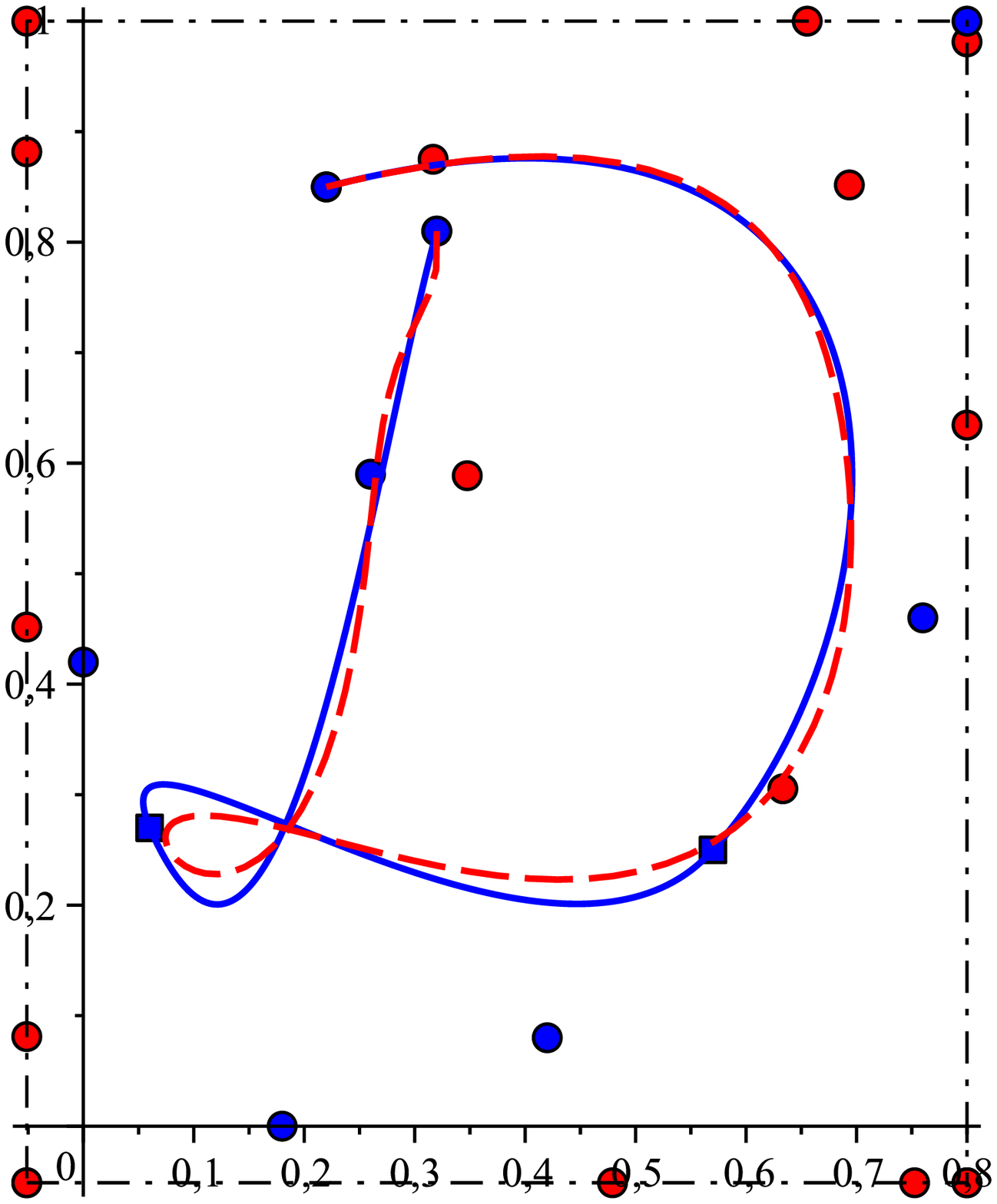}}
\subfloat[]{\label{figure5c}\includegraphics[width=0.33\textwidth]{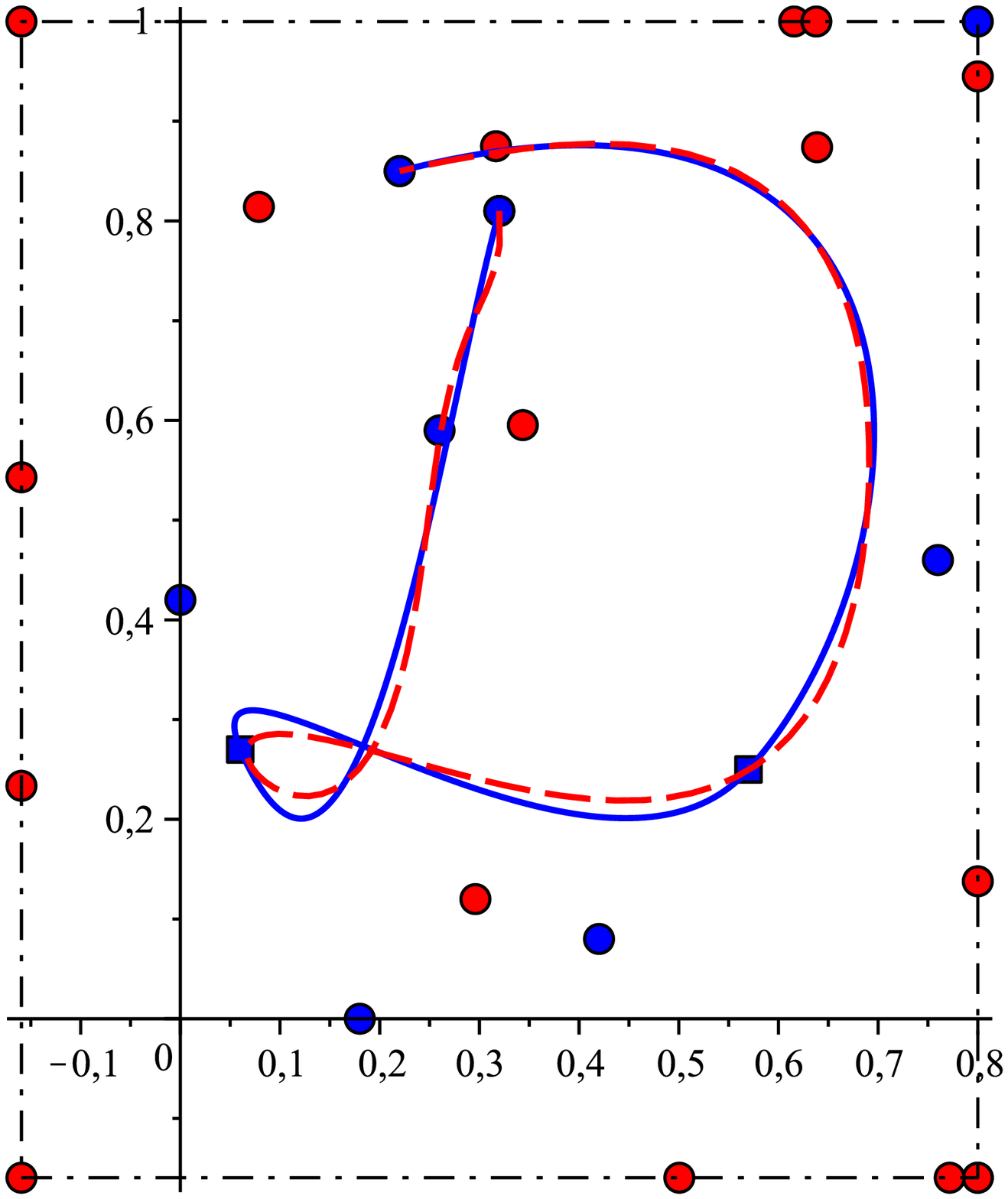}}
\end{tabular}
    \caption{Merging of three segments of the composite B\'{e}zier curve. The original composite curve (blue solid line with blue control points) and the merged curve (red dashed line with red control points) satisfying the following box constraints (black dotted-dashed frames): \eqref{Eq:DBox1} (see Figure~(a)), \eqref{Eq:DBox2} (see Figure~(b)) and \eqref{Eq:DBox3} (see Figure~(c)). Parameters: $m = 18$, $k = 1$, $l = 2$.}
\end{center}
\end{figure}

\end{remark}

\begin{remark}\label{R:Time}
To solve the box-constrained quadratic programming problem~\eqref{E:QP}, one can choose a method provided by a software library of a selected programming language or implement one of the algorithms given in~\cite{Fer98,Han90,Mor89}. For that reason, the running times strongly depend on the implementation of the selected method. However, regardless of the choice, the box constraints make Problem~\ref{P:Problem} more difficult to solve. Therefore, the running times of methods dealing with the new problem must be longer than in the case of the traditional merging. See the comparison given in Table~\ref{tab:table2}.

\begin{table}[H]
\captionsetup{margin=0pt, font={scriptsize}}
\centering
\ra{1.3}
\scalebox{1}{
\begin{tabular}{@{}clcccc@{}}
 \toprule & \phantom{a} &
Traditional merging & \phantom{a} &
\multicolumn{2}{c}{Problem~\ref{P:Problem}}\\
 \cmidrule{3-3} \cmidrule{5-6} & & Running times [ms] & & Running times [ms] & Iterations
\\ \midrule
Example~\ref{Ex:1} & & $29$ & & $224$ & $(15,17)$\\

Example~\ref{Ex:2} & & $50$ &  & $414$ & $(22,29)$\\

Example~\ref{Ex:3} & & $72$ & & $690$ & $(18,27)$\\
\hline
\end{tabular}}
\caption{Running times of the traditional and box-constrained merging of B\'ezier curves.}
\label{tab:table2}
\end{table}
\end{remark}

\section{Conclusions}\label{Sec:Conc}

The new approach to the problem of merging of B\'ezier curves is introduced. We propose constraints of the new type
and explain the purpose of those restrictions. A curve being the solution of Problem~\ref{P:Problem} is suitable for further modification
and applications. Moreover, the resulting convex hull is much smaller than the one obtained using
the traditional approach. Consequently, it can be helpful while solving some important problems.
These positive attributes make the new problem worth of consideration, despite the inevitably longer running times
and the larger approximation errors, which are also unavoidable. What is more, the comparison of the results
with the ones from~\cite{Gos15}, leads to a conclusion that in the case of the traditional merging, the defect
described in Section~\ref{Sec:motiv} is even more significant.

In the near future, the authors intend to study a more general version of Problem~\ref{P:Problem} with the geometric continuity
constraints instead of the conditions~\eqref{cont}. Furthermore, a different strategy of setting the restricted area could also improve the results.

\section*{Acknowledgments}
The authors are grateful to the referees for their remarks which helped to improve
the paper.


\end{document}